\documentclass[conference]{IEEEtran}
\IEEEoverridecommandlockouts

\usepackage[tbtags]{amsmath} 
\usepackage{verbatim} 
\usepackage{amsxtra}  
\usepackage{comment}
\usepackage{nicefrac}
\usepackage{bbm}

\usepackage{enumerate}

\usepackage{amsfonts}
\usepackage{color}

\usepackage{subfig}

\usepackage{wasysym}

\usepackage{cite}

\usepackage{amsthm}

\theoremstyle{definition}
\newtheorem*{defn*}{Definition}
\newtheorem*{scheme*}{Scheme}

\theoremstyle{remark}
\newtheorem{remark}{Remark}

\usepackage{fdsymbol}

\usepackage{tikz}
\usetikzlibrary{shapes,arrows,decorations.markings,positioning,calc}

\tikzstyle{plant} = [draw, fill=red!5, rectangle, 
    minimum height=3em, minimum width=6em]
\tikzstyle{block} = [draw, fill=blue!5, rectangle, 
    minimum height=3em, minimum width=18mm]
\tikzstyle{sum} = [draw, fill=yellow!10, circle, node distance=1cm]
\tikzstyle{coord} = [coordinate]
\tikzstyle{gain} = [draw, fill=red!5, regular polygon, regular polygon sides=3, shape border rotate=-90]
\tikzstyle{pinstyle} = [pin edge={to-,thick,black}]

\tikzstyle{BitPipe} = [thick, decoration={markings,mark=at position
   1 with {\arrow[semithick]{open triangle 60}}},
   double distance=1.4pt, shorten >= 5.5pt,
   preaction = {decorate},
   postaction = {draw,line width=1.4pt, white,shorten >= 4.5pt}]

\usetikzlibrary{fit,backgrounds}

\usepackage{epsfig, graphicx, psfrag}

\usepackage[mathcal]{euscript}
\usepackage[cal=boondox]{mathalfa}

\providecommand{\secref}[1]{Sec.~\ref{#1}}

\providecommand{\figref}[1]{Fig.~\ref{#1}}

\providecommand{\appref}[1]{App.~\ref{#1}}

\usepackage{refcount}

\DeclareMathOperator*{\argmin}{arg\, min}
\providecommand{\Exp}[1]{{\operatorname{exp} \left\{ #1 \right\}}}
\providecommand{\E}[1]{\ensuremath{\mathbb{E} \left[ #1 \right]}}

\newcommand{\Comment}[1]{}
\newcommand{\old}[1]{}
\newcommand{\rem}[1]{}

\providecommand{\ht}{{\hat{t}}}

\newcommand{\bY}{{\bf Y}}

\providecommand{\comment}[1]{}

\newcommand{\beqn}[1]{\begin{eqnarray}\label{#1}}
\newcommand{\eeqn}{\end{eqnarray}}
\newcommand{\beq}[1]{\begin{equation}\label{#1}}
\newcommand{\eeq}{\end{equation}}

\makeatletter
\newcommand{\vast}{\bBigg@{4}}
\newcommand{\Vast}{\bBigg@{5}}
\makeatother

\providecommand{\KL}[2]{{\bbD \left( #1 \middle\| #2 \right)}}

\providecommand{\bbD}{\mathbb{D}}

\providecommand{\E}[1]{\bbE \left[ #1 \right]}

\providecommand{\Exp}[1]{\ensuremath{{\operatorname{exp} \left\{ #1
\right\}}}}

\usepackage{comment}
\usepackage[draft=true]{hyperref}

\usepackage{graphicx,amsmath} 

\newcommand*{\bigsubseteq}{\mathbin{\scalebox{1.3}{\ensuremath{\subseteq}}}}

\newcommand*{\medcap}{\mathbin{\scalebox{1.3}{\ensuremath{\cup}}}}

\title{
Stochastic Chase Decoding for BMS Channels via  Rate Distortion Theory
}

\author{
\IEEEauthorblockN{
Amit Berman\IEEEauthorrefmark{1},
Ariel Doubchak\IEEEauthorrefmark{1},  
Uri Erez\IEEEauthorrefmark{2}\IEEEauthorrefmark{1},
Tal Philosof\IEEEauthorrefmark{1}
and 
Ilya Shapir\IEEEauthorrefmark{1}}
\IEEEauthorblockA{\IEEEauthorrefmark{1}Samsung Semiconductor Research and Development Center, Tel-Aviv, Israel, \\ \{amit.berman, ariel.d, tal.philosof, ilya.shapir\}@samsung.com.\\
\IEEEauthorrefmark{2}School of Electrical Engineering, Tel Aviv University, Israel, uri@eng.tau.ac.il
}
}

\begin{document}
\date{}
\maketitle

\begin{abstract}    
This work develops a rate-distortion-based approach to stochastic Chase decoding of algebraic codes over  binary memoryless
symmetric  (BMS) channels, replacing the heuristics traditionally used to determine flip probabilities with information-theoretically grounded flipping rules. In particular, we reinterpret stochastic Chase decoding as a random-coding construction for error-pattern covering codes.
Our approach builds on the framework of Nguyen et al., who introduced a rate-distortion formulation of multiple-attempt decoding for Reed-Solomon codes over nonbinary channels. In their formulation, erasure patterns are generated so as to align with, and thereby mask, hard-decision errors. We adapt this framework to the design of bit-flip probabilities for Chase decoding over BMS channels. This yields an explicit characterization of the asymptotically optimal bit-flipping rule, together with the expected list size required to ensure that the transmitted codeword appears in the decoding list with high probability. Moreover, for binary and quaternary symmetric channels, we demonstrate that the optimal bit-flipping rule, determined by exhaustive search,  closely matches the information-theoretic rule even at short block lengths.

\end{abstract}

\section{Introduction}

In recent years, there has been renewed  interest in practical decoding algorithms that attain near-maximum-likelihood (near-ML) performance for short- to medium-length codes in the context of Ultra-Reliable Low-Latency Communications (URLLC); 
see, e.g., Section~III of \cite{miao2024trends} for a survey. 
Similarly, near-ML decoding of constituent algebraic
codes is of interest in the decoding of generalized concatenated codes for flash memories; see, e.g., Section~VI of \cite{rajab2020soft}.

Because structured codes, and algebraic codes in particular, are highly competitive at short block lengths (see, e.g., \cite{van2016performance}), it is of considerable interest to develop near-ML decoding algorithms for this class of codes. In the setting of algebraic codes, such approaches have traditionally been studied within the framework  of soft-decision (SD) decoding.

There are several 
well-established SD decoding methods for algebraic linear block codes. Some are applicable to any such code, while others make use of the algebraic properties of the specific code. The former class includes ordered statistics decoding \cite{fossorier1995soft}, 
multiple-bases belief-propagation decoding, trellis-based maximum-likelihood (ML) decoders (applicable when the number of states is reasonable) \cite{wolf1978efficient}, 
and, more recently, neural net decoders \cite{nachmani2018deep}.


Approaches tailored to the algebraic properties of a code include the list decoders of \cite{koetter2003algebraic,wu2008new}, the ML decoder for BCH codes of \cite{vardy2002maximum}, as well as the family of Chase decoding methods. In the latter approach,
 originating in the work of Chase \cite{chase2003class}, multiple decoding attempts using a bounded-distance decoder (BDD) are employed. 
Chase decoding has numerous ramifications, see, e.g., \cite{kaneko2002efficient,kaneko2002improvement,fossorier2000chase,zhang2013efficient,kamiya2002algebraic,weber2004limited,shany2023fast}. 

SD approaches seek to improve decoding performance by exploiting reliability information from the received signal.
Early work focused primarily on achieving the asymptotic coding gain of ML decoding over AWGN channels, an appropriate metric when operating at rates far from capacity.

In the present work, however, we focus on higher-rate regimes. In particular, we consider the noise regime in which ML decoding attains a small error probability whereas BDD fails with high probability. Our goal is to design a stochastic Chase decoder that achieves a small error probability in this regime, which corresponds to end-of-life operation in flash memories \cite{rajab2020soft}.





Stochastic Chase decoding was introduced in  \cite{leroux2010stochastic}. 
A comparison of its performance against that of traditional Chase decoding for BCH codes over the binary-input AWGN (BI-AWGN) channel was empirically studied in
\cite{harada2015stochastic}. While these works demonstrated that stochastic Chase decoding can offer improved performance, both the underlying rationale and the flip probabilities used to generate the test patterns were heuristic.

Another line of relevant work is that of \cite{tokushige2003selection}, where test patterns were interpreted as a covering code. From this viewpoint, good covering codes are natural candidates for test-pattern generation. While constructing optimal covering codes is generally difficult, rate-distortion theory implies that nearly optimal covering codes can be obtained asymptotically via random coding. This perspective naturally leads to interpreting stochastic Chase decoding as a practical random-coding mechanism for generating asymptotically efficient test patterns.

The present work builds on Nguyen et al. \cite{nguyen2011multiple} 
who showed that 
the multiple decoding attempts approach can be formulated using rate–distortion (RD) theory by viewing it as an error pattern ``covering'' (lossy source coding) problem.\footnote{We use the terms ``multiple decoding attempts" and stochastic Chase decoding interchangeably, although the former may be viewed as more general since the code need not be randomly generated.}

The focus of \cite{nguyen2011multiple}, however, was on nonbinary channels and error-and-\emph{erasure} decoding, where the test patterns specify which symbols to erase. Their results are therefore directly applicable to Reed–Solomon codes. In contrast, the present work addresses binary memoryless
symmetric  (BMS) channels, where the test patterns specify which bits to flip, and derives the asymptotically optimal bit-flipping rule. Thus, the results are applicable to (binary) BCH codes where the appropriate distortion measure is Hamming distance rather than the metric employed in \cite{nguyen2011multiple}.

In summary, the goal of this work is to develop a principled rate-distortion-theoretic characterization of stochastic Chase decoding over BMS channels. In particular, we provide:
\begin{itemize}
\item a covering-code interpretation of stochastic Chase decoding,
\item a bit-flipping probability assignment rule that can be computed via a reverse water-filling formula,
\item an explicit characterization of the reliability cutoff, i.e., the channel reliability beyond which no flipping is performed,
\item an estimate of the number of error patterns that must be checked to ensure that the transmitted codeword appears in the list with high probability.

\end{itemize}


\begin{figure}[htbp]
    \centering
        \includegraphics[scale=0.12
        ]{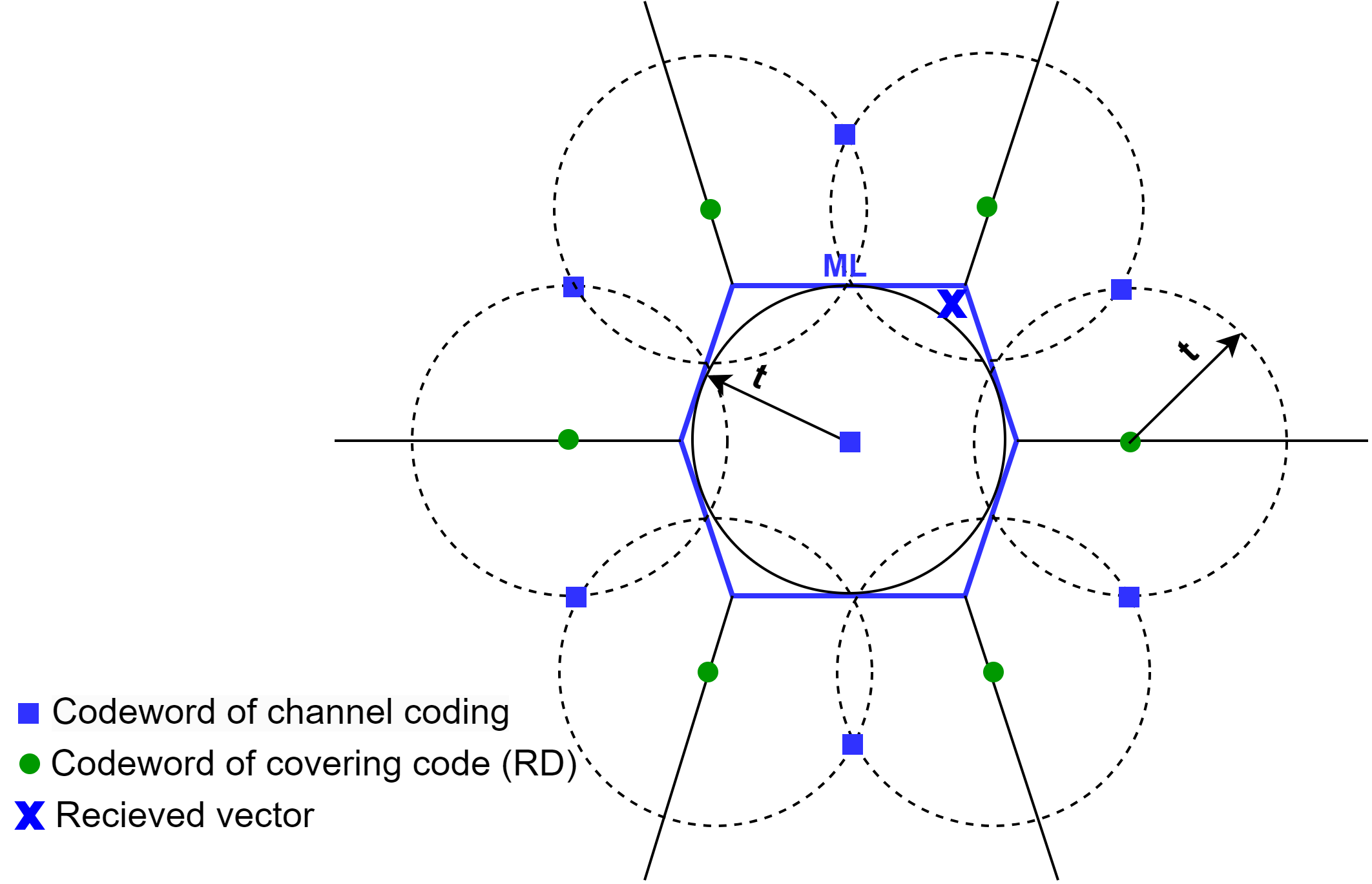}
        \caption{Geometric view of stochastic Chase decoding.
        }
        \label{fig:GV_no_P}
\end{figure}

\subsection{
Information-Theoretic Intuition}
Although Chase decoding is primarily used in the context of SD decoding, the results of the present work are most easily illustrated by first considering a binary symmetric channel (BSC), i.e., hard-decision (HD) decoding. The goal is to enable decoding beyond the code’s nominal error-correction radius.

For a generic binary linear code of rate $R=k/N$
and block length 
$N$, the worst-case complexity of ML decoding is roughly 
$2^{N \min(R,1-R)}$. 
Namely, $2^{NR}$ corresponds to a codeword domain search whereas $2^{N(1-R)}$ corresponds to an 
error pattern domain implementation, where one searches over the 
set of
 coset leaders (one per syndrome, via a table).
For a good code operating near capacity on a BSC with crossover probability 
$p$, we have 
$R \approx 1-H(p)$. Assuming $R>\nicefrac{1}{2}$, this translates to a search over roughly 
$2^{NH(p)}$ candidates.\footnote{Alternatively, $2^{NH(p)}$ is also 
the complexity of a sphere decoder 
that searches for a codeword within a typical noise sphere of radius approximately $np$ around the received vector.} 

Consider now a binary linear code that has an efficient BDD, allowing to correct all error patterns of weight no greater than 
\begin{align}
 t \triangleq \left\lfloor \frac{d_{\rm{min}}-1}{2} \right\rfloor.  \label{eq:t_def} 
\end{align}
This corresponds to the Hamming ball depicted in \figref{fig:GV_no_P},
the largest ball contained within the fundamental ML decoding region (shown as a hexagon).
Assuming a typical number of errors of approximately $Np$, over a BSC, a conventional Chase decoder requires adding to the received vector all error patterns of weight $Np-t$ in order to obtain a residual error pattern of weight at most $t$. The number of BDD attempts needed to ensure that at least one attempt succeeds is therefore equal to the cardinality of this sphere, i.e., one needs to search over a list with

\begin{align}
    {\rm list \ size} \approx  2^{N(H(p-t/N))}.
\label{eq:lossless_size}
\end{align}

Of course, each decoding trial has the complexity of BDD, but the overall complexity is nonetheless greatly reduced compared to generic ML or sphere decoding. 

A further reduction in complexity is possible by invoking RD theory in the spirit of \cite{nguyen2011multiple},  
treating the error sequence as an i.i.d.  ${\rm Ber}(p)$ (Bernoulli $p$) source, and the set of flip patterns as codewords of a lossy source (covering) codebook. 
Namely, the goal is to construct a codebook such that, with high probability, one can match an error pattern to a codeword whose difference (sum) has Hamming weight no greater than $t$.
By RD theory, assuming both $N$ and $t$ are large, and $\nicefrac{1}{2}>p>t/N$, a codebook/list with
\begin{align}
    {\rm list \ size} \approx 2^{N(H(p)-H(t/N))}
    \label{eq:rdf_size}
\end{align} 
should suffice, which is a substantial improvement over conventional Chase decoding.\footnote{This is due to the sub-additivity of the binary entropy function.} 
As noted in \cite{nguyen2011multiple}, the underlying insight of viewing pattern selection as a covering problem can be traced back to 
\cite{tokushige2003selection}.\footnote{We note that the syndrome checks in GRAND-type decoding \cite{duffy2019capacity} may be viewed as employing a BDD with $t=0$, in which case \eqref{eq:lossless_size} and \eqref{eq:rdf_size} coincide and reduce to the baseline complexity of $2^{NH(p)}$.
}


Similarly, any BMS channel may be viewed as a mixture of BSC channels with random state $S$, known to the receiver, dictating the crossover probability $p_s$.  In this case, as detailed in Section~\ref{sec:RD_for_chase}, the flip probability is a function of $s$, given by reverse water-filling. In the case where all states are active, the minimal size of the codebook will be roughly $2^{N(\mathbb{E}_S[H(p_S)]-H(t/N))}$. In the more general case, as will be established, the water level dictates the reliability threshold beyond which flipping is not beneficial. Consequently, for those states, no rate need be paid.




\section{Problem Formulation}
In the following, we introduce the channel model and  the decoding framework considered.

Assume a binary linear code of length $N$ is used over a  BMS channel.
We denote the channel binary input by $X_i \in \mathbb{F}_2$, $i=1,\ldots,N$. 

Any BMS channel is equivalent to a state-dependent BSC with i.i.d. state (drawn independent of the input) known to the receiver.
Specifically, in this representation the channel output is $Y_i=(\hat{X}^{\rm HD}_i,S_i)$ where $\hat{X}^{\rm HD}_i \in \mathbb{F}_2$ is the hard-decision and where $S_i \in \{ 1,\ldots,M\}$ is the state of the channel. We refer to the state as the \emph{reliability}  of the HD decision. In other words, given $S_i=j$, the channel from $X_i$ to $\hat{X}^{\rm HD}_i$ is a BSC with crossover probability $p_j \leq \nicefrac{1}{2}$:
$$
 E_i|S_i=j \sim \mathrm{Ber}(p_j),$$
where $E_i = X_i \oplus \hat{X}_i^{\rm HD}$. 

Thus, $i=1,\ldots,N$ is used as a time index and $j=1,\ldots,M$ is used as an index for  the channel state.
We further denote $\mathbf{p} = (p_1,\ldots,p_M)$ and
 $\mathbf{E}=(E_1,\ldots,E_N)$. The HD crossover probability is given by
\begin{align}
    p=\Pr(E_i=1)=\sum_j p_j \Pr(S_i=j).
    \label{eq:phd}
\end{align}





We refer to channel uses in which the state is $j$, as belonging to \emph{reliability class} $j$ and denote by $N_j$ the number of transmitted bits within class $j$. Thus, $N_j=\sum_{i=1}^N \mathbf{1}\{S_i =j\}$
and $N=\sum_{j=1}^M N_j$. 
We further denote $\mathbf{N}=(N_1,\ldots,N_M)$.

For the most part we will condition on the vector $\mathbf{S}=(S_1,\ldots,S_N)$, which induces also $\mathbf{N}$, so that the statistics by which  $S$ is drawn will not play a role. 
Nonetheless, in the limit of block length $N \rightarrow \infty$, the composition of the vector $\mathbf{N}$ will tend to the dominant type, and thus is dictated by the distribution of $S$. This limit is exploited when we analyze the extension of the results to the case of a 
BI-AWGN channel.

\section{Rate-Distortion Framework for Multiple Attempts decoding: Noise Level Above BDD Capability}
\label{sec:RD_for_chase}

We  adapt the approach of \cite{nguyen2011multiple} to the considered channel model.
We assume the availability of an efficient BDD for up to $t$
HD errors \eqref{eq:t_def} and 
that  $L$ BDD attempts are performed to produce a list of up to $L$ candidate codewords, from which we choose the one with the highest likelihood.

We define $\mathcal{E}_{L}$ as the event that the transmitted codeword does not appear in the list and denote the error event of an ML decoder by $\mathcal{E}_{\rm{ML}}$. 

Note that the event in which the transmitted codeword is included in the list but does not have the highest likelihood is a subset of  $\mathcal{E}_{\mathrm{ML}}$. This follows since an ML decoder will also fail when the transmitted codeword is not the one with the highest likelihood. 
The error event of the proposed decoder, denoted by $\mathcal{E}$, therefore satisfies
\begin{align}
    \mathcal{E} \bigsubseteq  \mathcal{E}_{\rm{ML}} \medcap \mathcal{E}_{L}.
\end{align}

Applying the union bound yields 
\begin{align}
    \Pr(\mathcal{E}) \leq \Pr(\mathcal{E}_{\rm{ML}}) + \Pr(\mathcal{E}_{L}).
\end{align}
We aim to minimize the probability of $\mathcal{E}_{L}$, which represents the performance loss relative to the \emph{optimal} ML decoder. Thus, we henceforth focus our attention on the event $\mathcal{E}_{L}$.

We are interested in the regime where while the ML error probability is small, BDD likely fails, i.e., the HD crossover probability $p$ as defined in (\ref{eq:phd}), satisfies $p>t/N$.
We wish to design (a sequence of) codebooks with $L$ test error patterns  such that 
$\Pr(\mathcal{E}_L)$ vanishes as $N \rightarrow \infty$. 
More specifically, the goal is to find the minimal (exponential) growth rate of $L$ (as a function of $N$) such that this holds. RD theory answers precisely this question as detailed next.

To that end, note that a BDD attempt is successful iff the error test pattern
$\hat{\mathbf{E}}$ is such that  
\begin{align}
|\hat{\mathbf{E}} \oplus \mathbf{E}| \leq t  
\label{eq:wrap}
\end{align}
where $| \cdot |$ denotes Hamming weight and $t$ is defined in \eqref{eq:t_def}. 

Setting $D=t/N$, this may equivalently be written as requiring that the average Hamming distortion between the two vectors satisfies
\begin{align}
d(\mathbf{E},\hat{\mathbf{E}}) =\frac{1}{N}\sum_{i=1}^N \mathbf{1}\{E_i \ne \hat E_i\} \leq D.    
\label{eq:distortion}
\end{align}
Thus, $\mathcal{E}_L$ is the event that none of the $L$ patterns satisfies \eqref{eq:distortion}. 

We can rewrite the distortion as 
\begin{align}
d(\mathbf{E},\hat{\mathbf{E}}) &=\frac{1}{N}\sum_{j=1}^M \sum_{i=1}^N \mathbf{1}\{ \{E_i \ne \hat E_i\} \cap \{S_i=j\} \}  \\
&=\frac{1}{N}\sum_{j=1}^M  N_j d(\mathbf{E}_{|j},\hat{\mathbf{E}}_{|j}) 
,    
\label{eq:distortion2}
\end{align}
where $\mathbf{E}_{|j}$ denotes the sub-vector of $\mathbf{E}$ obtained from retaining only coordinates for which $S_i=j$.

It follows that one can interpret the problem through the lens of lossy source coding where the list of test error patterns is the reconstruction codebook. 
Thus, it is amenable to standard RD analysis. 
Specifically, since the distortion measure is additive, the problem reduces to 
finding the RDF of $M$ parallel sources subject to per-source distortion constraints $D_j$
 satisfying
\[
\frac{1}{N}\sum_{j=1}^M N_j D_j \;\le\; D =\frac{t}{N}, \qquad 0 \le D_j \le p_j.
\]

We next recall the relevant rate-distortion function.

\subsection{RDF for Parallel Bernoulli Sources}
Consider first the case of a single reliability class, i.e., $M=1$ and  denote $p_1=p$ where $0 \leq p \leq \nicefrac{1}{2}$.
Recall that for a $\mathrm{Ber}(p)$ source, under  Hamming distortion, the RDF is given by
\begin{align} 
R_p(D) = \max \{0, H(p) - H(D) \}
\label{eq:RDF}
\end{align}
where $H(\cdot)$ is the binary entropy function \cite{coverthomas}, and the reverse (test) channel is a BSC with crossover probability $D$ with $\rm{Ber}(q)$ input where
\begin{align}
q= \max \left\{0, \frac{p-D}{1-2D} \right\}.    
\label{eq:optimal_q_0}
\end{align}

For the discrete BMS model considered, we may partition the block of total length $N$, into $M$ sub-blocks. Sub-block $j$ is of length $N_j$, with error sequence generated according to an i.i.d. Bernoulli $p_j$ distribution. Thus, the problem amounts to source coding for a set of $M$ independent parallel Bernoulli sources. 

Assuming that $N_j$ is large for all $j$, we may again invoke rate-distortion theory. 
The sum RDF is given by
\begin{align}
R_{\mathbf{p}}(D)=\sum_{j=1}^M  \frac{N_j}{N} \Big[ H(p_j) - H(D_j^\star) \Big],\label{eq:Rp}
\end{align}
where the $\{D_j^\star\}$ solve the minimization:
\begin{align}
    \min_{\{D_j\}} \sum_{j=1}^M  \frac{N_j}{N} \Big[ H(p_j) - H(D_j)\Big]\label{eq:D_opt}
\end{align}subject to
\begin{align}   
\sum_{j=1}^M \frac{N_j}{N} D_j \le D, 
\qquad 0 \le D_j \le p_j.
\end{align}
This minimization leads to the reverse water-filling solution as recalled next.

Specifically, applying the KKT conditions, the optimal distortion allocation
\begin{align}
D_j^\star = \min\!\left\{ \nu, \, p_j \right\}, 
\qquad \sum_{j=1}^M  \frac{N_j}{N}D_j^\star = D,    
\label{eq:water1}
\end{align}
with a common water level $\nu$ and where $p_j,q_j,D_j^*$ satisfy the  reverse channel relation:
\[
p_j = q_j (1-D_j^\star) + (1-q_j) D_j^\star \;=\; q_j \otimes D_j^\star,
\]
where $\otimes$ denotes binary convolution.
In other words, we set 
\begin{align}
q_j = \max \left\{0, \frac{p_j-\nu}{1-2\nu} \right\},\; j=1,\ldots,M.    
\label{eq:optimal_q}
\end{align}


\subsection{Application to Stochastic Chase Decoding}\label{sec:RD_Stochastic_chase}
The RD solution suggests a disciplined method to set the flip probability in stochastic Chase decoding by identifying the list of randomly generated  error patterns with the RDF codebook as detailed next.

{
\textit{Generation of test patterns codebook based on RDF:}
\begin{itemize}
\item[$\smallblackcircle$] Set $D = t/N$.
\item[$\smallblackcircle$] Find the optimal water level $\nu$ by solving \eqref{eq:D_opt}.
\item[$\smallblackcircle$] Set $D_j$ according to \eqref{eq:water1}.
\item[$\smallblackcircle$] Set the flip probabilities $q_j$ according to \eqref{eq:optimal_q}.
\item[$\smallblackcircle$] Set the codebook size to $L \approx 2^{N R_p(D)}$, where $R_p(D)$ is given in \eqref{eq:Rp}.
\item[$\smallblackcircle$] 
Draw $L$ independent codewords with independent (but not identically distributed) entries according to 
\begin{align}
\hat{E}_i \mid S_i = j \sim \mathrm{Ber}(q_j),
\label{eq:testchannel}
\end{align} 
$i=1,\ldots,N$, based on the reliability class assignment $S_i$, to obtain the test error codebook $\{\hat{\mathbf{E}}^{(\ell)}\}_{\ell=1}^{L}$.
\item[$\smallblackcircle$] For $\ell=1,\ldots,L$, apply BDD to decode $\bY \oplus \hat{\mathbf{E}}^{(\ell)}$.
\item[$\smallblackcircle$] Choose among the decoded codewords (if any), the one with the highest likelihood. 
\end{itemize}

}

\subsection{Example: BMS Channel with two Reliability Classes}

\begin{figure}[htbp]
    \centering
    \vspace{-0.0cm}    
    \subfloat[Channel probabilities $p_1$ and $p_2$.]{\includegraphics[width=0.48\columnwidth, height=4cm]{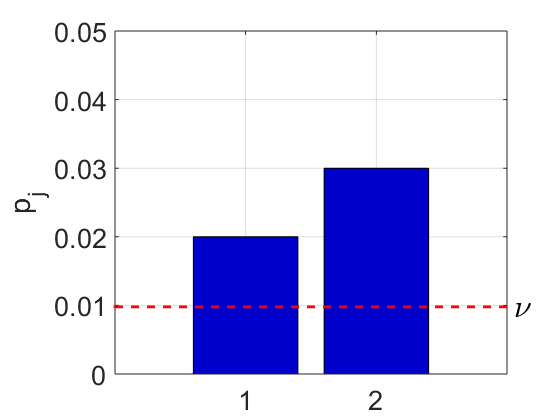} 
    \label{fig:2SD_Pr}}
    \subfloat[Flip probabilities $q_1$ and $q_2$.]    {\includegraphics[width=0.48\columnwidth, height=4cm]
    {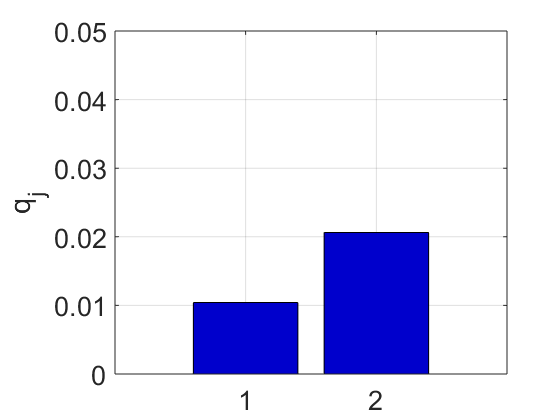}
    \label{fig:2SD_Pflip}}
    \caption{Example of BMS channel with two reliability classes.}
    \label{fig:2SD_Pr_Pflip}
\end{figure}

Consider a BMS channel with two reliability classes, i.e., $M = 2$. The two crossover probabilities are $p_1 = 0.02$ and  $p_2 = 0.03$ as depicted in \figref{fig:2SD_Pr}.
We set the block length to $N = 511$ and assume a BDD with $t = 5$. We further assume that the channel output vector has $N_1 = 477$ symbols in the higher reliability class and $N_2 = 40$ in the lower reliability  class.

Following the steps in \secref{sec:RD_Stochastic_chase} yields the common water level $\nu$, determined via \eqref{eq:D_opt}, which is shown in \figref{fig:2SD_Pr} by the red dashed line. The corresponding flip probabilities $q_1$ and $q_2$, obtained from \eqref{eq:optimal_q}, are depicted in \figref{fig:2SD_Pflip}.


\subsection{Example: BI-AWGN Channel}
We may (informally) extend the results to the BI-AWGN channel by viewing it as a limiting case with $M \rightarrow \infty$, as outlined next.

Let the channel output be 
\begin{align}
    \tilde{Y}_i=\tilde{X}_i+Z_i,
\end{align}
where $\tilde{X}_i=(-1)^{X_i}$ and $Z_i \sim N\left(0,\sigma^2 \right)$.
The state of the channel is now a continuous reliability value, which we parametrize via the absolute value of the log-likelihood ratio (LLR). Namely,
$$S_i=\left|\log\left( \frac{f_{\tilde{Y}|X}(\tilde{Y}_i|x=0)}{f_{\tilde{Y}|X}(\tilde{Y}_i|x=1)}\right) \right|=\frac{2r_i}{\sigma^2},$$
where $r_i=|\tilde{Y}_i|$.
Thus, the channel output can be viewed as the pair  $Y_i=(\hat{X}^{\rm HD}_i,S_i)$. 
Given that $S_i=\ell_i$, the channel from $X_i$ to $\hat{X}^{\rm HD}_i$ is a BSC with crossover probability 
$p(\ell_i)$, where 
\begin{align}
    p(\ell) 
    &=\frac{1}{1+e^{{\ell}}}.
\end{align}
The reverse water-filling solution now becomes
 \begin{align}
 \frac{1}{N}\sum_{i=1}^N  D^\star_i = D,
 \label{eq:awgn_water1}   
 \end{align}
 where
 \begin{align}
 D^\star_i = \min\left\{ \nu, \, p(\ell_i) \right\},
 \label{eq:awgn_water2}    
 \end{align}
 and where $\ell_i$ 
denotes the realization of the channel state $S_i$ at time $i$.
The flip probability $q_i$ is computed via \eqref{eq:optimal_q} substituting $p_j$ with $p(\ell_i)$.



\begin{figure}[htbp]
    \centering
    \vspace{-0.4cm}    
    \subfloat[$p(\ell_i=2r_i/\sigma^2)$ vs. sorted channel index.]{\includegraphics[width=0.48\columnwidth, height=4.5cm]{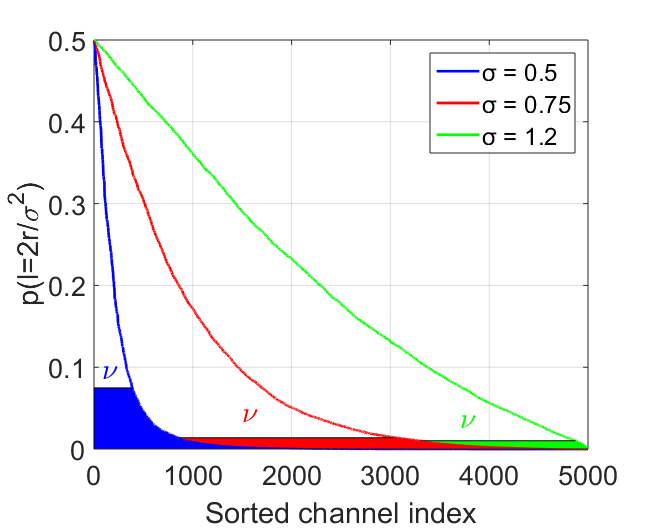} 
    \label{fig:BI-AWGN_Pr}}
    \subfloat[$q(\ell)$ vs. LLR.]    {\includegraphics[width=0.48\columnwidth, height=4.5cm]
    {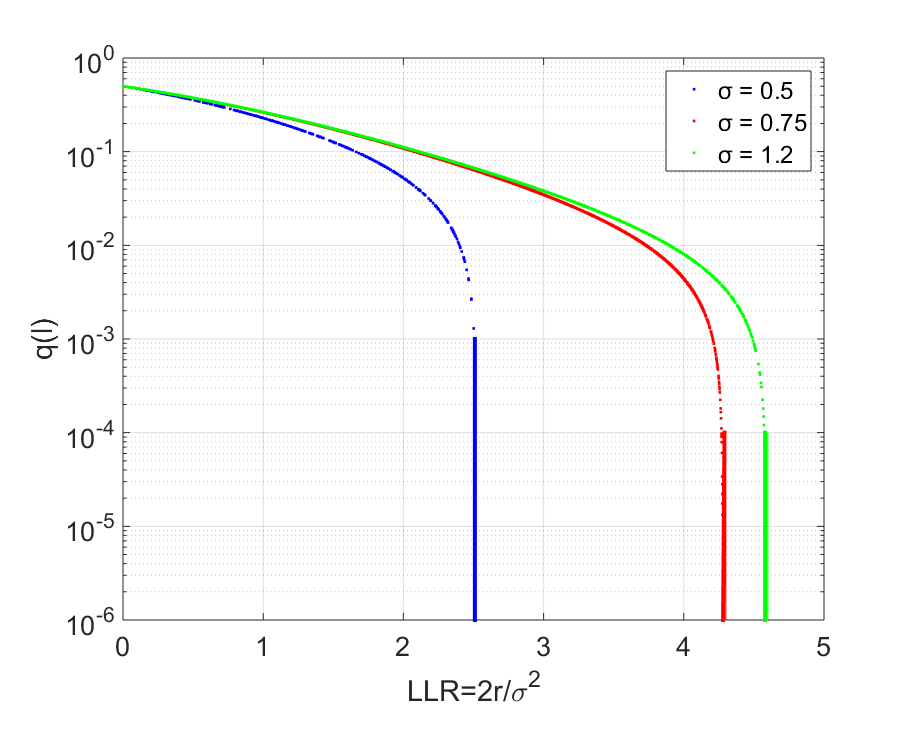}
    \label{fig:BI-AWGN_Pflip}}
    \caption{Reverse water-filling distortion allocation and flip probability rule for BI-AWGN channel at different SNR values.}
    \label{fig:BI-AWGN_Pr_Pflip}
\end{figure}

\figref{fig:BI-AWGN_Pr} depicts the probability $p(\ell=2r_i/\sigma^2)$ as a function of $i$ for $\sigma = 0.5, 0.75, 1.2$, where we order the channel index in increasing reliability. 
The colored regions represent the summands in \eqref{eq:awgn_water1}, which sum to $ND$ (irrespective of $\sigma$). 
\figref{fig:BI-AWGN_Pflip} illustrates the  corresponding flip probability rule $q(\ell)$ as a function of $\mathrm{LLR}$ for each $\sigma$. Note that the rule changes solely due to the change in the water level $\nu$ in \eqref{eq:optimal_q}. The latter also prescribes the LLR threshold beyond which no flipping should be performed, i.e., those indices $i$ for which $p(\ell_i)<\nu$.

\begin{remark}
    
In the limit of $N \rightarrow \infty$, the water-level equation \eqref{eq:awgn_water1} becomes the integral
\begin{align}
\int_{r=0}^\infty  f(r) \min\left\{ \nu, \, p\left(\ell=\frac{2 r}{\sigma^2} \right) \right\}  dr= D,    
\label{eq:gaussian_water}
\end{align}
where
\begin{align*}  
f(r)=\frac{1}{\sqrt{2 \pi \sigma^2}} \left[ \exp\left(-\frac{(r+1)^2}{2 \sigma^2} \right)+\exp\left(-\frac{(r-1)^2}{2 \sigma^2} \right)\right],r \geq 0,
\end{align*}
is the density of the absolute value of the channel output $\tilde{Y}_i$.
\end{remark}

\section{Numerical Results}



In \appref{app:opt_P_flip}, we derive the failure probability  $\Pr(\mathcal{E}_L|\mathbf{N})$ given in \eqref{eq:pe_given_N} for $L$  BDD attempts, conditioned on the state  vector $\mathbf{N}$. The  flip probabilities $\tilde{\mathbf{q}} \triangleq (\tilde{q}_{1},\ldots,\tilde{q}_{M})$ are selected such that the failure probability is minimized \eqref{eq:opt_pf_given_N}. 


In this section, we numerically investigate the RDF-based predictions and compare them with the optimal solution. We do so for the BSC ($M=1$) and for a BMS channel with $M=2$ reliability classes. The number of BDD attempts is taken to be that dictated by the RD solution, i.e., $L=2^{NR_p(D)}$ 
where $R_p(D)$ is given in \eqref{eq:Rp}.
We compare the RDF-derived flip probabilities $q_j$, derived in Section~\ref{sec:RD_for_chase},   to the optimal flip bit probabilities $\tilde{q}_j$, i.e., those solving 
\eqref{eq:opt_pf_given_N}.

Note the computational complexity of the optimal solution is $O(N^M)$ and is prohibitive unless $M$ is small and $N$ is moderate.

\begin{figure}[htbp]
    \centering
    \includegraphics[width= \columnwidth]{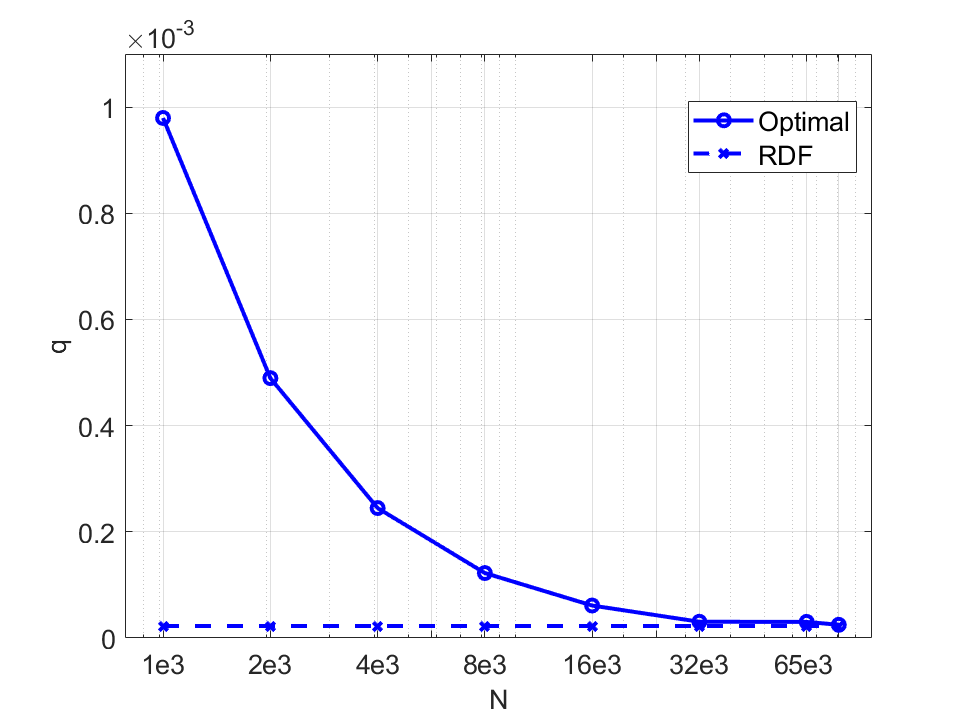} 
     \caption{Flip probabilities vs. block length for $M=1$  with the specified parameters.}
    \label{fig:HD_pflip_vs_N}
    \vspace{-0.5cm}    
\end{figure}

We first demonstrate the results for the case of $M=1$, i.e., a BSC channel with crossover probability $p$. \figref{fig:HD_pflip_vs_N} depicts the flip probability of the test patterns, for the RDF-based solution, as well as that of the optimal solution, the latter as a function of  block length $N$. Specifically, we evaluate the performance for fixed $t/N$ where $t/N = 9.8 \cdot 10^{-4}$ and a BSC with crossover probabilities $p = 10^{-3}$. 
As expected, we observe that, as $N$ increases, the optimal solution converges to the RDF solution.

\begin{figure}[htbp]
    \centering
    \includegraphics[width= \columnwidth]{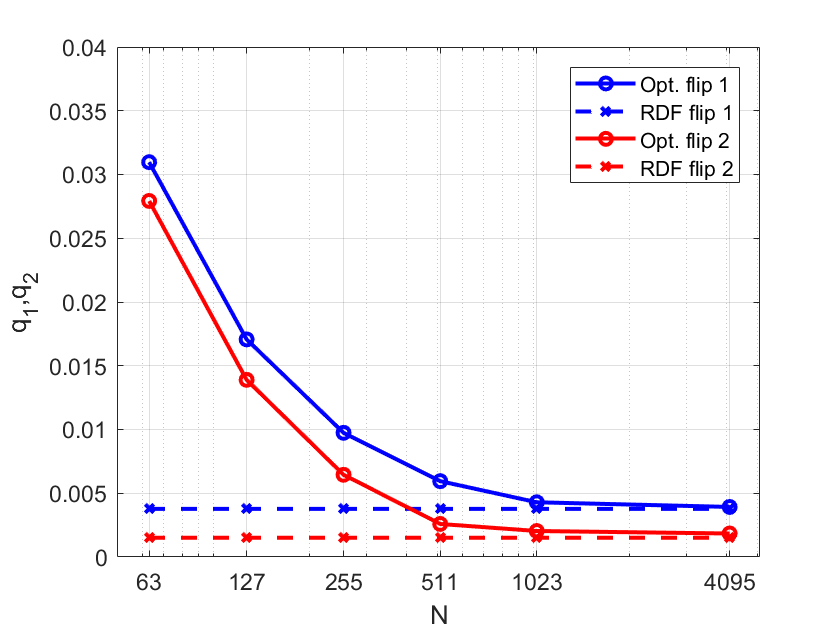}
    \caption{Flip probabilities vs. block length for $M=2$ with the specified parameters.}
    \label{fig:2SD_pflip_vs_N}
    \vspace{-0.5cm}    
\end{figure}

We next demonstrate the results for $M=2$ reliability classes. Specifically,
\figref{fig:2SD_pflip_vs_N}  depicts the flip probabilities $\tilde{q}_{1}$ and $\tilde{q}_{2}$ for the two reliability classes as a function of $N$ where $t/N = 0.079$ is held fixed, the crossover probabilities are $p_{1} = 0.083$ and $p_{2} = 0.081$, and where we take $N_{1} = 0.4N$ and $N_{2} = 0.6N$. We observe that, indeed, the optimal solution approaches the RD solution as $N$ increases. 


\section{Discussion}
\label{sec:below}

The analysis carried out is useful when the noise level is such that the typical number of errors is higher than the BDD correction capability $t$. In this regime, rate-distortion analysis yields a non-zero rate (list size) and a non-zero flip probability for at least one reliability class. 

Chase decoding has been traditionally used, however, also in the regime where the typical number of errors is smaller than $t$. In this regime, the ``first order" analysis does not offer useful insights or guidelines since the RDF vanishes and a single BDD attempt suffices to ensure vanishing error probability as the block length grows.

Nonetheless, it is possible to obtain meaningful results for this regime as well by carrying out a more refined rate-distortion analysis as proposed and carried  out for Reed Solomon codes in  \cite{nguyen2011multiple}. 
The relevant tool is the rate-distortion exponent (RDE) and its application to the scenario of interest in this paper is left for future work.
We further note that while random coding is asymptotically optimal, there remains considerable room for improvement in the design of finite-blocklength test-pattern sets and covering codes.

\appendices 

\section{Optimal Flip Probabilities for Stochastic Chase Decoding over a BMS Channel}\label{app:opt_P_flip}

Assume a BMS channel with $M$ reliability classes and a BDD with error-correction  capability $t$. For each trial, a flip pattern is applied, where each bit in the pattern is drawn independently according to ${\rm Ber}(\tilde{q}_j)$, $j = 1, \ldots, M$, according to its reliability class. The error probability of a single trial corresponds to the event that the total number of errors remaining after  bit flipping exceeds the decoder’s correction capability $t$. In \secref{sec:RD_for_chase}, this error event for $L$ test patterns was denoted by $\mathcal{E}_L$, i.e., the event  in which the transmitted codeword does not appear in the list.

We further denote by $n_{e,j}$ and $n_{c,j}$ the number of errors that remain after bit flipping within the error pattern and outside the error pattern, respectively, for reliability class $j$.

The error probability for a single flip pattern ($L = 1$), given specific values of $\mathbf{N}_e$ and $\mathbf{N}_c$, where $\mathbf{N}_c \triangleq \mathbf{N} - \mathbf{N}_e$, is given by:
\begin{align}
    &\Pr(\mathcal{E}_1 | \mathbf{N}_{e},\mathbf{N}_{c}) = 
    \sum_{\sum_{j=1}^M n_{e,j}+n_{c,j}  > t} \prod_{k=1}^M\binom{N_{e,k}}{n_{e,k}}\binom{N_{c,k}}{n_{c,k}}
    \nonumber \\ 
    &\qquad\cdot \tilde{q}_k^{N_{e,k}+n_{c,k}-n_{e,k}} (1-\tilde{q}_k)^{N_{c,k} - n_{c,k} +n_{e,k}},  \label{eq:pe_given_Ne_Nc}
\end{align}
where $\mathbf{N}_e = (N_{e,1},\ldots,N_{e,M})$ and $\mathbf{N}_c = (N_{c,1},\ldots,N_{c,M})$. Thus, the error probability for $L$ flip patterns drawn independently is given by 
\begin{align}
    \Pr(\mathcal{E}_L | \mathbf{N}_{e},\mathbf{N}_{c}) = \left[\Pr(\mathcal{E}_1 | \mathbf{N}_{e},\mathbf{N}_{c})\right]^L.\label{eq:L_err}
\end{align}

We are now ready to express the error probability for $L$ trials given the reliability class composition vector $\mathbf{N}$:
\begin{align}
    &\Pr(\mathcal{E}_L|\mathbf{N}) = \sum_{N_e > t} \Pr(\mathcal{E}_L,N_e|\mathbf{N}) \\ 
    & = \sum_{N_e > t} \sum_{\sum_{j=1}^m N_{e,j} = N_e} \Pr(\mathcal{E}_L,\mathbf{N}_e|\mathbf{N}) \\ 
    & = \sum_{N_e > t} \sum_{\sum_{j=1}^m N_{e,j} = N_e} \Pr(\mathbf{N}_e|\mathbf{N}) \Pr(\mathcal{E}_L|\mathbf{N}_e,\mathbf{N}_c)\label{eq:EQN_100}\\    
    &=\sum_{N_e > t} \sum_{\sum_{j=1}^M N_{e,j}=N_e} \prod_{k=1}^M\binom{N_k}{N_{e,k}}
    p_k^{N_{e,k}}(1-p_k)^{N_{c,k}} \nonumber \\
    &\qquad \cdot
    \left[\Pr(\mathcal{E}_1 | \mathbf{N}_{e},\mathbf{N}_{c})\right]^L,\label{eq:pe_given_N}
\end{align}
where \eqref{eq:EQN_100} follows from the fact that
$\Pr(\mathcal{E}_L \mid \mathbf{N}, \mathbf{N}_e) = \Pr(\mathcal{E}_L \mid \mathbf{N}_e, \mathbf{N}_c)$. The last equality \eqref{eq:pe_given_N} follows from \eqref{eq:L_err}, where the probability $\Pr(\mathcal{E}_1 \mid \mathbf{N}_e, \mathbf{N}_c)$ is given in \eqref{eq:pe_given_Ne_Nc}.

Thus, the optimal flip probabilities of the test patterns $\tilde{q}_{j}$, $j = 1, \ldots, M$, can be expressed as:
\begin{align}
    \tilde{\mathbf{q}}^*(\mathbf{N}) = \argmin_{\tilde{\mathbf{q}}} \Pr(\mathcal{E}_L|\mathbf{N}),\label{eq:opt_pf_given_N}
\end{align}
where $\Pr(\mathcal{E}_L|\mathbf{N})$ is given in \eqref{eq:pe_given_N}, and $\tilde{\mathbf{q}} \triangleq (\tilde{q_{1}},\ldots,\tilde{q}_{M})$.

\bibliographystyle{IEEEtran}
\bibliography{mybib}


\end{document}